\documentclass[12pt]{article}
\usepackage{graphicx,amsmath,amssymb,color,bm}

\begin{document}

\centerline {\Large\textbf {Field-created diverse quantizations in phosphorenes
}}

%\centerline {\Large \textbf {}}\vskip0.6 truecm

\centerline{Jhao-Ying Wu$^{1}$, Szu-Chao Chen$^{2}$, and Godfrey Gumbs$^{3}$}

\centerline{$^{1}$Center of General Studies, National Kaohsiung Marine University, Kaohsiung, Taiwan 811}
\centerline{$^{2}$Department of Physics, National Cheng Kung University,
Tainan, Taiwan 701}
\centerline{$^{3}$Department of Physics and Astronomy, Hunter College at the City University of New York,\\ \small }
\centerline{695 Park Avenue, New York, New York 10065, USA}

\begin{abstract}

 Electronic properties of few-layer phosphorenes are investigated by the generalized tight-binding model. They are greatly diversified by the electric and magnetic fields ($E_z$ and $B_z$). The $E_z$-induced gap transition, Dirac cones, oscillatory bands and critical points are present in bilayer system, but absent in monolayer one. The diverse magnetic quantization phenomena cover the coexistent two subgroups of Landau levels, the uniform and non-uniform energy spacings, and the crossing and anti-crossing behaviors. Specifically, the wavefunctions exhibit the dramatic changes between the well-behaved and  multi-mode oscillations. The feature-rich energy spectra are reveled in density of states as .many special structures which could be verified from scanning tunneling spectroscopy.

\vskip0.6 truecm


\end{abstract}

\newpage

\bigskip

\centerline {\textbf {I. INTRODUCTION}}%

The two-dimensional layered systems, with the nano-scaled thickness and the unique geometric symmetries, have stirred a lot of experimental and theoretical studies \cite{Xu2013,Sheneve2013,Wang2015}. They have been successfully synthesized by various experimental methods, such as, graphene \cite{Aissa2015}, silicene \cite{Takagi:2015}, germanene \cite{Derivaz:2015}, tinene \cite{Zhu2015}, and transition metal oxides \cite{Wang2012}. Such 2D systems are very suitable for studying the novel physical, chemical and material phenomena. Specifically, few-layer phosphorenes are recently produced by using the mechanical cleavage approach \cite{PLi2014,PLiu2014}, liquid exfoliation \cite{PBrent2014,PYasaei2015,Pkang2015}, and mineralizer-assisted short-way transport reaction \cite{PLange2007,PNilges2008,PKopf2014}. These systems inherently have energy gaps of $~1.5-2.0$ eV \cite{gap1,gap}, as identified from optical measurements \cite{PLiu2014,Zhang2014}. Such gaps are higher than that (${\sim\,0.2-0.3 eV}$ eV) of bulk system \cite{Li2014,Han2014}, and  they are
in sharp contrast with the zero or narrow gaps of 2D group-IV systems \cite{Balendhran2015}. Transport measurements show that the phosporene-based field-effect transistor exhibits an on/off ratio of 105 and a carrier mobility at room temperature as high as 103 cm$^{2}/$V.s \cite{PLi}. Furthermore, monolayer phosphorene displays the unusual energy spectra and quantum Hall effect due to the magnetic quantization \cite{Likai2016}. Few-layer phosphorenes are expected to have high potentials in the next-generation electronic devices and optical devices \cite{PLi,TLow}. This work is focused on how to create the diverse quantization phenomena in monolayer and bilayer phosphorenes by tuning a composite magnetic and electric field (${{\bf B}=B_z\hat z}$ $\&$ ${{\bf E}=E_z\hat z}$).

Each phosphorene layer possesses a puckered structure, mainly owing to the $sp^{3}$ hybridization of ($3s,3p_{x},3p_{y},3p_{z}$) orbitals. The deformed
hexagonal lattice on x-y plane is quite different from honeycomb lattices of group-IV systems \cite{PRudenko}. This unique geometric structure fully dominates the low-lying energy bands, in which they are highly anisotropic in dispersion relations of energy versus wave vector, e.g., the linear and parabolic dispersions near $E_{F}$, respectively, along $\widehat{k_{x}}$ and $\widehat{k_{y}}$ directions \cite{PRudenko}. The anisotropic behaviors are clearly revealed in other physical properties, as verified by recent measurements on optical spectra and transport properties \cite{PLi,TLow}. This is a unique advantage of phosphorene in comparison with Mo$S_2$-related and semiconductors. The unusual anisotropy could be utilized in the design of unconventional thermoelectric devices. For example, the thermal gradient and the potential difference are applied in two orthogonal directions, leading to one with the higher thermal conductivity and another with the larger electrical conductivity \cite{LingX2015}. Moreover, this intrinsic property will greatly diversify the quantization phenomena.

%Transport measurements indicate that there is a structural anisotropy of phosphorus which is reflected through a carrier mobility in the range of $\approx 10^3$cm$^2$V$^{-1}$s$^{-1}$.\cite{phos1,phos2,phos3,phos4,phos5,phos6,phos7}

The low-lying electronic structure is easily tuned by the external electric and magnetic fields. A uniform perpendicular electric field can create the monotonous increase of energy gap with strength in monolayer phosphorene. Specifically, bilayer phosphorene presents the drastic changes of energy bands and becomes a gapless system after the critical electric field ($E_{z,c}$) \cite{PDolui,PLiu2015}. There exist rich energy dispersions during the variation of $E_z$, including the parabolic bands, the graphene-like  Dirac-cone structure, and the oscillatory bands. The unusual transition comes from the strong competitive or cooperative relations among the intralayer and the interlayer atomic interactions, and the Coulomb potentials. These will be directly reflected in the diverse magnetic quantization phenomena. The generalized tight-binding (TB) model is further developed to explore the essential properties in detail \cite{CYLin2015}. The Hamiltonin is built from the tight-binding functions on the distinct sublattices and layers, in which all the interactions and external fields are taken into account simultaneously. This method can deal with the the magnetic quantization of electronic states even in the presence of complicated geometric structures and external fields \cite{CYLin2015,ChenSC2016}.

The dispersionless Landau levels (LLs) come from the magnetic quantization of neighboring electronic states. The main features are investigated for monolayer and bilayer
phosporenes in a composite electric and magnetic field, especially for the ${B_z}$- and $E_{z}$-dependent energy spectra and the spatial distributions of quantum modes. The generalized tight-binding model is suitable for studying the competitive quantization due to the multi-constant energy loops and the coexistent extreme and saddle points in the energy-wave-vector space. This study shows that the LL spectra exhibit the monotonous or non-monotonous dependences, and the non-crossing, crossing or anti-crossing behaviors.
Furthermore, there are two kinds of LLs according to the well-behaved and perturbed distribution modes. The anti-crossing spectra will be clearly illustrated by the obvious changes in the mixing modes. Specifically, two distinct subgroups of valence (conduction) LLs near the Fermi level are identified from the distinguishable localization centers. They are never observed in the other 2D systems up to now. The unusual energy spectra are directly revealed in the special structures of density of states (DOS). They could be verified from  experimental measurements of scanning tunneling spectroscopy (STS).

\bigskip
\bigskip
\centerline {\textbf {II. METHODS}}%
\bigskip
\bigskip

Monolayer phosphorene, with a puckered honeycomb structure, has a rectangular unit cell (black lines in Fig. 1(a)). There are four phosphorus atoms, in which half of them are located at the lower or higher (A and B) sublattices. The similar structures are revealed in few-layer systems, e.g, bilayer phosphorene in Fig. 1(b). The low-lying energy bands are dominated by the atomic interactions of 3$p_{z}$ orbitals \cite{PRudenko}. The few-layer Hamiltonian is characterized by
\begin{equation}
H=\sum_{i=1,l}^{4}(\varepsilon_{i}^{l}+U_{i}^{l})c_{i}^{l}c_{i}^{\dag\,l}+\sum_{\langle i,j \rangle, l}t^{ll}_{ij}c_{i}^{l}c_{j}^{\dag\,l}+\sum_{\langle i,j \rangle, l \neq l^{\prime}}t^{\prime\,ll^\prime}_{ij}c_{i}^{l}c_{j}^{\dag\,l^{\prime}}.
\end{equation}
$\varepsilon_{i}^{l}$ is zero in monolayer; that of few-layer system is the layer- and sublattice-dependent site energy due to the chemical environment. $U_{i}^{l}$ is Coulomb potential energy induced by an electric field. $c_{i}^{l}$ ($c^{\dag\,l^{\prime}}_{j}$) is annihilation (creation) operator.  $t^{ll}_{ij}$ and $t^{\prime\,ll^\prime}_{ij}$ are, respectively, the intralayer and interlayer hopping integrals, and the effective interactions used in the calculations cover the forth and fifth neighboring atoms. These hopping parameters are adopted from Ref. \cite{PRudenko}.

Monolayer and bilayer phosphorene are assumed to exist in a uniform perpendicular magnetic field. The magnetic flux through a unit rectangle is $\Phi= a_{1}a_{2}B_z$. $a_{1}=3.27$ ${\AA}$ and $a_{2}=4.43$ ${\AA}$ are lattice constants (Fig. 1). The vector potential, $\vec{A}=(B_z x)\hat{y}$ can create an extra magnetic Peierls phase of $exp\{i[\frac{2\pi}{\phi_{0}}\int \vec{A}\cdot d\vec{r}] \}$, leading to a new period along $\hat x$ and thus an enlarged rectangular unit cell with ${4R_B= 4\phi_0\,/\Phi}$ atoms in monolayer system. $\phi_{0}$ (=h/e=4.1$\times 10^{-15}$ T$\centerdot m^{2}$) is the magnetic flux quantum; ${\phi_0\,/\Phi}$ is chosen to be an integer. The reduced first Brilloun zone has an area of ${4\pi^2\,/a_1a_2R_B}$. For bilayer phosphorene, the magnetic Hamiltonian matrix is a very huge 8$R_{B_{0}}$$\times$8$R_{B_{0}}$ Hermitian matrix within the experimental field strength, e.g., the dimension of 16800 at ${B_0=30}$ T. The eigenvalues and eigenstates could be computed very efficiently by the exact diagonalization method.

%\begin{equation*}
%H=\sum_{\langle ij\rangle \langle ll^{\prime }\rangle }-t_{ij}^{ll^{\prime
%}}C_{il}^{\dag}C_{jl^{\prime }},
%\end{equation*}%
%where $t_{ij}^{ll^{\prime }}$ represents the five intralayer (Fig. 13(a)) and four interlayer hopping integrals (Fig. 13(b); details in Ref. \cite{PRudenko}). For monolayer system, the magnetic Hamiltonian is a $4R_{B}\times 4R_{B}$ Hermitian matrix.

\bigskip
\bigskip
\centerline {\textbf {III. RESULTS AND DISCUSSION}}%
\bigskip
\bigskip

The special lattice structure and multiple hopping integrals are responsible for the rich energy bands. Monolayer phospoorene has a direct gap of ${E_g\sim\,1.6}$ eV near the $\Gamma$ point (Fig. 2(a)), while group-IV systems present zero or narrow gaps at the K point \cite{Lu2006}. Along $\Gamma$X and $\Gamma$Y directions, energy dispersions are approximately linear and parabolic, respectively. The effective mass of the former is much lighter than that of the latter, being associated with the preferred chemical bonding along $\hat x$ \cite{Qihang2015}. The conduction and valence bands are asymmetric about ${E_F=0}$. They, respectively, arise from the linearly symmetric and anti-symmetric superpositions of TB functions on the two sublattices (A and B). This simple relation of the same layer is modified in bilayer phosphorene, mainly owing to the finite site energies (the first term in Eq. (1)). Furthermore, the layer-dependent TB functions make the distinct contributions the two pairs of energy bands, as shown in Fig. 2(c). The smaller band gap of ${E_g\sim\,1.01}$ eV is reduced by the interlayer atomic interactions.

A perpendicular electric field can greatly diversify electronic properties. $E_g$ of monolayer system grows monotonously with the increasing field strength (Figs. 2(a) and 2(b)). As for bilayer system, the first pair of energy bands approaches to $E_F$ (blue curves in Figs. 2(c))-2(f)), while the opposite is true for the second pair (red curves; $E_z$ in unit of V/$\AA$). The parabolic bands of the former lead to zero gap near the $\Gamma$ point  at the critical field of ${E_{z,c}\,=0.3}$ (Fig. 2(f)). With the further increase of field strength, their energy dispersions present the dramatic changes (Fig. 2(g)). Along $\Gamma$Y and $\Gamma$X (${\hat k_x}$ and ${\hat k_y}$), there exist the linearly intersecting bands and the oscillatory bands, respectively. Two splitting Dirac-cone structures  are situated at the right- and left-hand sides of the $\Gamma$ point (along ${+\hat k_y}$ and ${-\hat\,k_y}$ in Fig. 2(h)). Furthermore, the extreme points are just the $\Gamma$ point, accompanied with two saddle points at the opposite $k_x's$. All the critical points and the constant-energy loops in the energy-wave-vector space will dominate the main features of LL spectra. Specifically, the Coulomb potential energy differences can create the significant probability transfer between two layers. For example, four TB functions might be extremely non-comparable for a sufficient high field strength. This will play critical roles in the unusual LL wavefunctions.

%The up-layer (down-layer) states may penetrate into the valence (conduction) band along the linear energy dispersion when $\Gamma$X-gap is opened in $E_{z}>E_{c}$ (Fig. 2(h)). This is mainly dominated by the increased electric potentials. Oppositely, there exists a strong interaction between the valence and conduction bands near the saddle points, i.e., the energy bands are severely hybridized with the almost equal weights of charge densities on two layers. This demonstrates the strong competition relation between the electric potential and the interlayer atomic interactions along $\Gamma$X direction. The feature-rich energy dispersion and charge density distributions, arising from the anisotropic lattice structure, would directly reflect in the magnetic quantization.

The highly anisotropic energy dispersions create the unique dependence of LL energies on the quantum number ($n^{c,v}$; discussed later) and the magnetic filed strength, as shown in Fig. 3. Each LL is four-fold degenerate for each (${k_x,k_y}$) state because of the spin degree and the mirror symmetry about z-axis. The $(k_x=0,k_y=0)$ state in the reduced Brillouin zone is chosen for a systematic study. In monolayer and bilayer phosphorenes, the low-lying LL energies cannot be well described by a linear relation ${n^{c,v}B_z}$ (the dashed pink lines), especially for the higher energy and field strength. This is different from the square-root dependence in monolayer graphene \cite{JHHo2008}, and the linear dependence in AB-stacked bilayer graphene \cite{Lai2008} and MoS$_{2}$ \cite{HoY2014}. There are two groups of valence and conduction LLs in bilayer phosphorene. Both of them cross each other frequently, in which the main differences lie in the initiated energies and the LL spacings. Specifically,  the initial LL energy corresponds to that of the extreme point in zero-$B_Z$ energy band. The intragroup and intergroup anticrossings, as revealed in intrinsic ABC-stacked graphene \cite{CYLin2015}, are absent, since all the well-behaved LLs are quantized from the monotonous band structure in the energy-wave-vector space (Figs. 2(a) and 2(c))

%The first- and second-groups wavefunctions, respectively, correspond to the in-phase and out-of-phase subenvelope functions on different sublayers (Fig. 4).

%The quantized LLs are characterized by the subenvelope functions on the different subplanes and sublayers. They are localized at the 1/2 and 2/2 positions of the enlarged unit cell, corresponding to the magnetic quantization at the $\Gamma$ point. \textbf{The two localization centers correspond to the linearly anti-symmetric and symmetric superpositions of subenvelope functions on the two layers, respectively (at $k_{x}=k_{y}=0$).} The $3p_{z}$-orbit quantization, with spin degree, is four-fold degenerate for each $(k_{x},k_{y})$ state. This is in sharp contrast to the eight-fold degeneracy in the group-IV systems, or the double degeneracy of the spin- and valley-dependent LLs in MoS$_2$. The LL degeneracy depends on the number of equivalent valleys and the existence of inversion symmetry ($z\rightarrow-z$ and $x\rightarrow-x$). If the Landau wavefunctions are well-behaved spatial distributions, the number of zero points of the oscillations can be used to determine the quantum number of each LL; otherwise, the LLs are undefined.

The main features of LLs are dramatically changed by the electric field for ${E_z\ge\,E_{z,c}}$. The LL spectrum could be divided into three regimes according to the energy ranges of the distinct band dispersions, e.g., $E_z$=0.32 in Figs. 4(a) and 4(b). (I), (II) and (III), respectively, correspond to the Dirac cone (green), the inner and outer parabolic bands (between the saddle and extreme points; blue and pink curves), and the parabolic one (below or above the extreme point; pink). For the valence states in (I), the zero-energy LLs are formed at the Fermi level, clearly indicating the magnetic quantization initiated from the $E_z$-induced Dirac point (the extreme point). State degeneracy of the low-lying LLs is twice than others, mainly owing to the splitting Dirac-cone structures with the almost same energies (Figs. 2(g) and 2(h)). The LL energy spacings quickly decline in the increase of quantum number. Specifically, the LLs in (II) present the abnormal sequence, since they come from the strong competition of magnetic quantization in the two distinct constant-energy loops. The LL spectrum becomes well-behaved in (III), and its energy spacing is almost uniform. This directly reflects the normal quantization of monotonous parabolic band.

The feature-rich LL spectrum could be fully understood from the Landau wavefunctions. The subenvelope functions are distributed on the different sublattices and layers. They are localized near the 1/2 and 2/2 positions of the enlarged unit cell, being related to the magnetic quantization at the $\Gamma$ point. The similar localization behaviors occur around  two centers, so the 1/2 position is sufficient for a model study. In general, the identical oscillation modes are revealed on the two sublattices of the same layer. The number of zero points in the dominant amplitude distribution is the quantum number. There are the coexistent main and side modes for the low-lying LL states at ${E_z\ge\,E_{z,c}}$, a feature due to the cooperation of the mutli-hopping integrals, the Coulomb potential energies, and the magnetic field. The subenvelope functions in (I) are simultaneously localized at the left- and right-hand sides of the 1/2 position (Figs. 4(c) and 4(d)), in which the two degenerate states exhibit the same weights on the distinct sublattices and layers (light and heavy green curves). They are associated with the magnetic quantization of two neighboring Dirac cones. Such LL states are regarded as the first subgroup of valence LLs. When state energies are gradually away from the Dirac points, the zero points in the spatial oscillations increase. The two separate localization centers will merge together, reducing the LL degeneracy to fourfold at ${E^v<-}$0.034 eV. The zero-point numbers of the ${l_A^1}$ and ${l_A^2}$ components grow quickly (pink curves) when entering into (II). The strong oscillation modes of  many zero points  clearly demonstrate that such LLs come from the quantized states of the larger constant-energy loop. On the other hand, the smaller constant-energy loop  can create the second subgroup of LLs (blue curves). The initial LLs at ${-0.053}$ eV, with the dominant zeroth mode in the ${l_A^1}$ component, correspond to the magnetic quantization near the $\Gamma$ point. The competitive quantization between two distinct loops leads to the abnormal LL sequence and spatial oscillations in (II); that is, there exist the crossings and anti-crossings of two subgroups (Fig. 5(a)). The subenvelope functions in (I) and (II) do not present a single-mode oscillation, and their components on the two layers might quite differ from each other. But for (III), they are all well-behaved oscillation modes arising from the monotonous parabolic dispersion, i.e., they are identical to those of a quantized oscillator \cite{Lai2008}. Specifically, the LL wavefuntions exhibit the same oscillation modes on the two distinct layers. In short, two subgroups of LLs are the magnetically quantized states near the Dirac and $\Gamma$ points; their strong competition induces the unusual quantization phenomena.

The LL energy spectra, as shown in Figs. 5(a) and 5(b), are greatly diversified by the external fields. The valence and conduction LLs exhibit the similar $B_z$- and $E_z$-dependent spectra. The former in (I) have the well-behaved $B_z$-dependence when the magnetic field is below 20 T (Fig. 5(a)). Their energies could be fitted by the square-root relation ($\sqrt{n^v_D B_z}$). This dependence is similar to that of monolayer graphene with the linear Dirac cone \cite{JHHo2008}. However, the two entangled LLs, which arise from the anti-crossing of two subgroups, appear at  higher magnetic fields. As for the valence LLs in (II), their spectrum presents the coexistent two subgroups under the non-monotonous relations. Furthermore, the ${n_D^v}$ and ${n_{\Gamma}^v}$ LLs have the opposite ${B_z}$-dependences.  The crossings and anti-crossings occur continuously in the alternate form. This abnormal $B_z$-dependent spectrum is never observed in other 2D materials; that is, the unusual  LL spectrum  of two competitive subgroups is absent in other layered systems. The anti-crossings clearly illustrate that such LLs are composed of the multi-oscillation modes (Fig. 6). With the deeper state energies, the LL spectrum changes into the monotonous $B_z$-dependence, directly reflecting the parabolic energy dispersion. All the LLs in (III) belong to the single-mode oscillations with many zero points. The complicated LL spectra are also achieved by tuning the electric field. At ${E_z>E_{z,c}}$ (Fig. 5(b)), there exist the frequent crossings and anti-crossings related to the two subgroups of LLs. The diverse LL spectra will be obviously revealed in  DOSs as the special structures, so that they could be directly identified from the STS measurements.

The anti-crossing spectra originate from the Landau states with the multi-zero points. To prevent the mixed LLs crossing each other, both of them have certain identical  oscillation modes. The drastic changes of the wave functions during the anticrossing processe between the fourth LLs in the first and the second subgroups are illustrated in Figs. 6(a) and 6(b) (blue and pink curves). The anticrossing region occurs in the range of $\sim$27.6-31.5 T, where the most serious hybridization of the two levels is present at its center. At the initial field strength, both ${n_{D}^v=3}$ and ${n_{\Gamma}^v=3}$ LLs have the main mode of three zero points in the ${l_A^1}$ component. Furthermore, the former possesses the side modes with five, seven and nine zero points. The latter presents the similar side modes in the increase of $B_z$. The side modes grow, but the main mode declines. Such oscillation modes are seriously mixed at the anti-crossing center ($B_z$=29.6 T). And then, the mutli-oscillation modes gradually become the behaved ones. The similar transformation of oscillation modes are revealed in the ${l^2_{A}}$ component, in which the main and side modes of the two anticrossing LLs, respectively, correspond to 10 and (4,6,8,12) zero points. The frequent LL anti-crossings occur in the $B_z$- and $E_z$-dependent energy spectra (Figs. 5(a) and 5(b)), since the external fields, and the intralayer and interlayer atomic interactions strongly compete or cooperate with one another.

%%%%%%%%%%%%%%%%%%%%%%%%%%%%%%%%%%%%%%%%%%%%%%%%%%%%%%%%%%%%%%%%%%%%%%%%%%%%%%%%%%%%%%%%%%%%%%

The main features of energy bands and LL spectra are directly reflected in DOS. At zero or weak fields, the band-edge states of parabolic bands create the gap-dependent shoulder structures, e.g., $E_z$=0.29 in Fig. 7(a) by the blue curve. The initial structures are replaced by a valley-like structure due to the deformed Dirac cone (Fig. 2(g)), when the gap transition happens at $E_{z}=E_{c}$ (Fig. 7(b)). The symmetric peaks in the logarithmically divergent form, corresponding to the saddle points in energy bands, come to exist at the both sides of this valley. The special structures are more obvious at $E_{z}>E_{c}$, especially for the extended valley structure (Figs. 7(c) and 7(d)). Furthermore, there exist the shoulder structures arising from the extreme $\Gamma$ points. The prominent peaks and shoulders are gradually  away from the Fermi in the further increase of $E_z$. The $E_z$-induced drastic changes in energy bands could be verified from the STS measurements on the energy gap and the valley-like, shoulder and peak structures \cite{Song2010,Jung2011}.

%The existence of logarithmically divergent peaks is an evident characteristic of the $E_{z}$-induced band overlap. The range of the linear valley shape structure and the frequency difference between the logarithmically divergent peak and the step structure are increased by the increment of $E_{z}$.

The magnetic field induces a lot of delta-function-like peaks. The height and spacing of peaks, respectively, reflect state degeneracy and energy dispersion of
$B_z$=0. At $E_{z}<E_{c}$, the low-frequency DOS peaks have the uniform height of the four-fold degeneracy and the almost same spacing (red curve in Fig. 7(a)), mainly owing to the quantization of parabolic band (Fig. 2(f)). But for ${E_z\ge\,E_{z,c}}$, the unusual features appear in the energy ranges of (I) and (II) (Figs. 7(b)-7(d)).  There are one pair of peaks centered about the Fermi level  at ${E_z=E_{z,c}}$ (Fig. 7(b)). With the increase of $E_z$,  a very prominent peak, with the eight-fold degeneracy, is  revealed at $E=0$ (Figs. 7(c) and 7(d)). The similar peaks, which come from the quantized Dirac cone,  could survive at stronger electric fields (Fig. 7(d)). The double-peak structures at higher/deeper energies are due to the two anti-crossing LLs. Apparently, all the low-lying peaks present the highly non-uniform spacings. The STS measurements on the main features of low-energy  LL peaks could provide the useful informations about the diverse magnetic quantizations.

Phosphorenes are in sharp contrast with graphenes in electronic properties, such as, the field-dependent band structures and LLs.  For monolayer systems, the former and the latter, respectively, present the middle and zero gaps associated with the parabolic and linear bands. The Dirac cone of graphene can create the square-root dependence in the $B_z$-dependent LL energy spectrum. Each LL is eight-fold degenerate because of the hexagonal symmetry (or two equivalent
valleys) \cite{Lai2008}. The AA- and AB-stacked bilayer graphenes are semimetals with band overlaps. The AA stacking could be regarded as the superposition of two monolayer graphenes in magnetic quantization; that is, this system has the well-behaved energy spectra and LL wavefunctions. However, the electric field in the AB stacking leads to an energy gap and the valley-split LLs. The LL degeneracy is reduced to half under the destruction of mirror symmetry about $z=0$ plane. Furthermore, each splitting LL subgroup exhibit the anti-crossing behavior. However, the coexistent magnetic quantization, which originates from the Dirac-cone structure and the two-constant loops in one valence (conduction) energy band, is absent in graphene systems. The lattice symmetries and the intralayer and interlayer atomic interactions are responsible for the critical differences.

\bigskip
\bigskip
\centerline {\textbf {IV. Concluding Remarks}}%
\bigskip
\bigskip

The generalized tight-binding model is developed to elucidate the electronic properties
of monolayer and bilayer phosphorenes in the external fields. The feature-rich characteristics include the $E_z$-induced drastic changes in band structures and the $B_z$-created diverse magnetic quantizations. The spatial distributions of the subenvelope functions are critical in illustrating the main features of LLs, some of which, such as the quantum number and subgroup classification, can be determined by the number of zero points and the dominant sublattices. This method provides an approach to describing other main-stream 2D materials under various fields. For example, there exist important differences between phosphorene and group-IV systems in quantization phenomena \cite{ChenSC2016}. Furthermore, it could combine with the single- and many-particle theories to explore the essential physical properties, such as, magneto-optical and Coulomb excitations \cite{Ho2010,Wu2011}.

A single-layer phosphorene only exhibits the monotonous dependence on $E_z$ and $B_z$ in terms of energy spectra and wavefunctions. The electric and magnetic fields can create the diverse phenomena in bilayer system, such as, the gap transition, the coexistent linear, oscillatory and parabolic bands, the two subgroups of LLs, the uniform and non-uniform LL energy spacings, and the frequent crossings and anti-crossings. The subenvelope functions present the dramatic changes between the well-behaved and multi-mode oscillations during the hybridization of two mixed LLs.
The main features of energy bands and LLs are reflected in DOS as various structures, including valleys, shoulders, and logarithmic and delta-function-like peaks. The number, form, height and energy of LL peaks near the Fermi level are closely related to the magnetic quantization arising from the Dirac cone and the two constant-energy loops, e.g., a stronger peak at E=0 and the double-peak structures.
The STS measurements on the low-lying special structures are useful in understanding the competitive or cooperative relations among the external fields, and the intralayer and interlayer atomic interactions.

\bigskip

\bigskip

\centerline {\textbf {ACKNOWLEDGMENT}}%

\bigskip

\bigskip

\noindent \textit{Acknowledgments.} This work was supported by the MOST of Taiwan, under Grant No. MOST 105-2112-M-022-001.

\newpage

\begin{figure}
\centering
\includegraphics[width=1\textwidth]{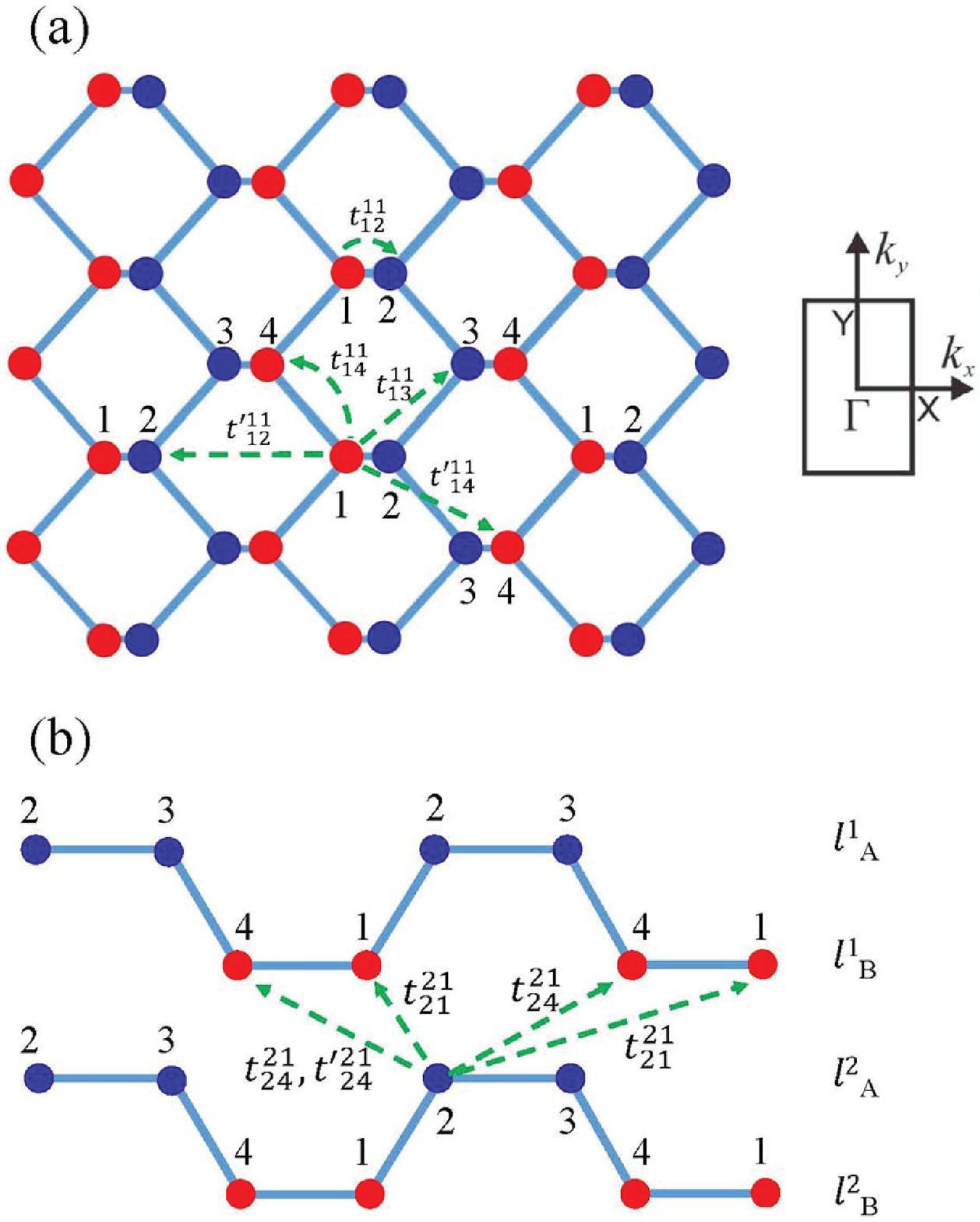}
\caption{Geometric structures of (a) monolayer and (b) bilayer phosphorenes, respectively, under the top and side views with various atomic interactions. Also shown is the first Brillouin zone.}
\label{FIG:1}
\end{figure}

\begin{figure}
\centering
\includegraphics[width=1\textwidth]{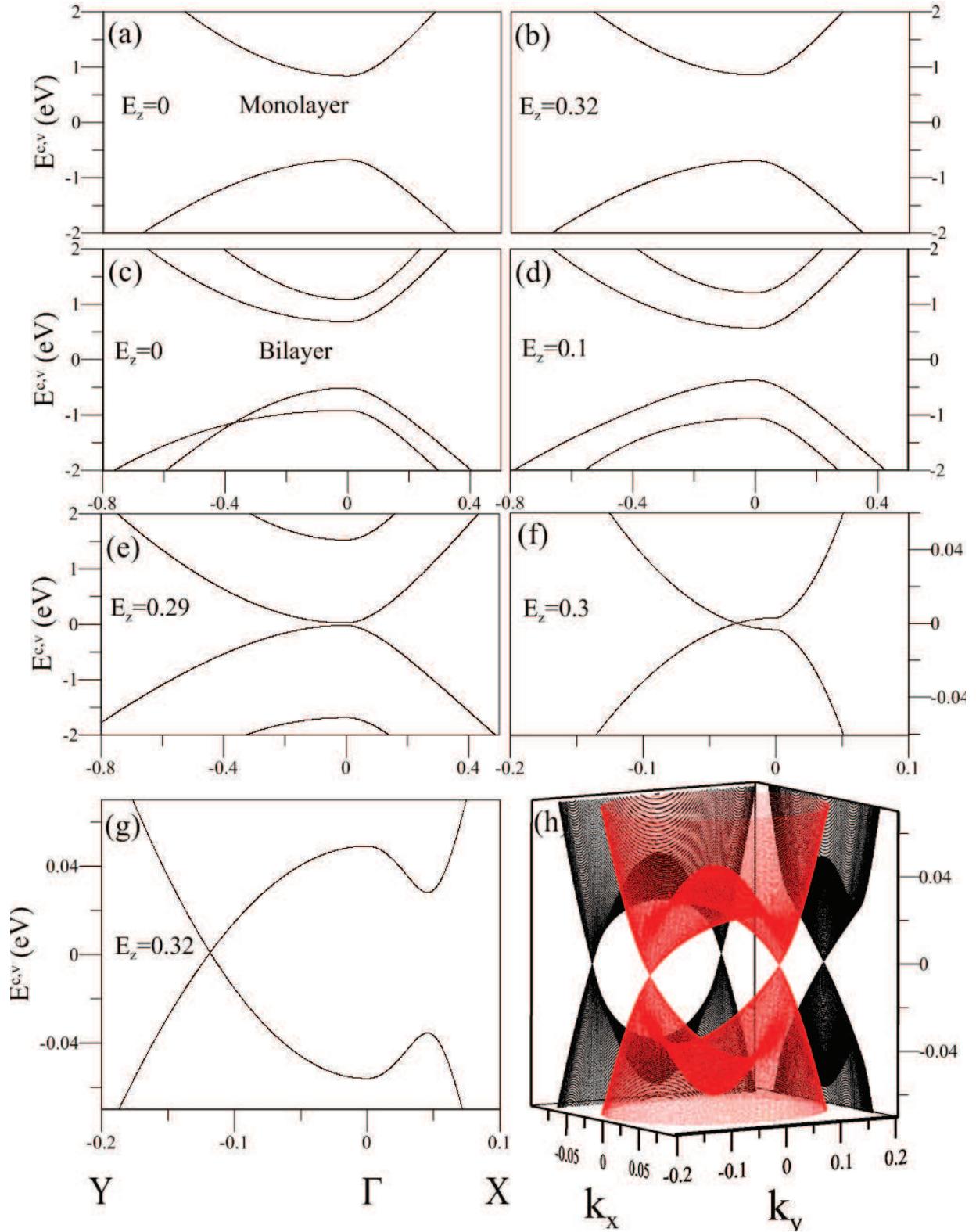}
\caption{The $E_z$-induced changes of electronic structures for monolayer phosphorene in (a)-(b) and bilayer system in (c)-(g). At $E_z$=0.32, the diverse band dispersions in energy-wave-vector space is shown in (h).}
\label{FIG:2}
\end{figure}

\begin{figure}
\centering
\includegraphics[width=1\textwidth]{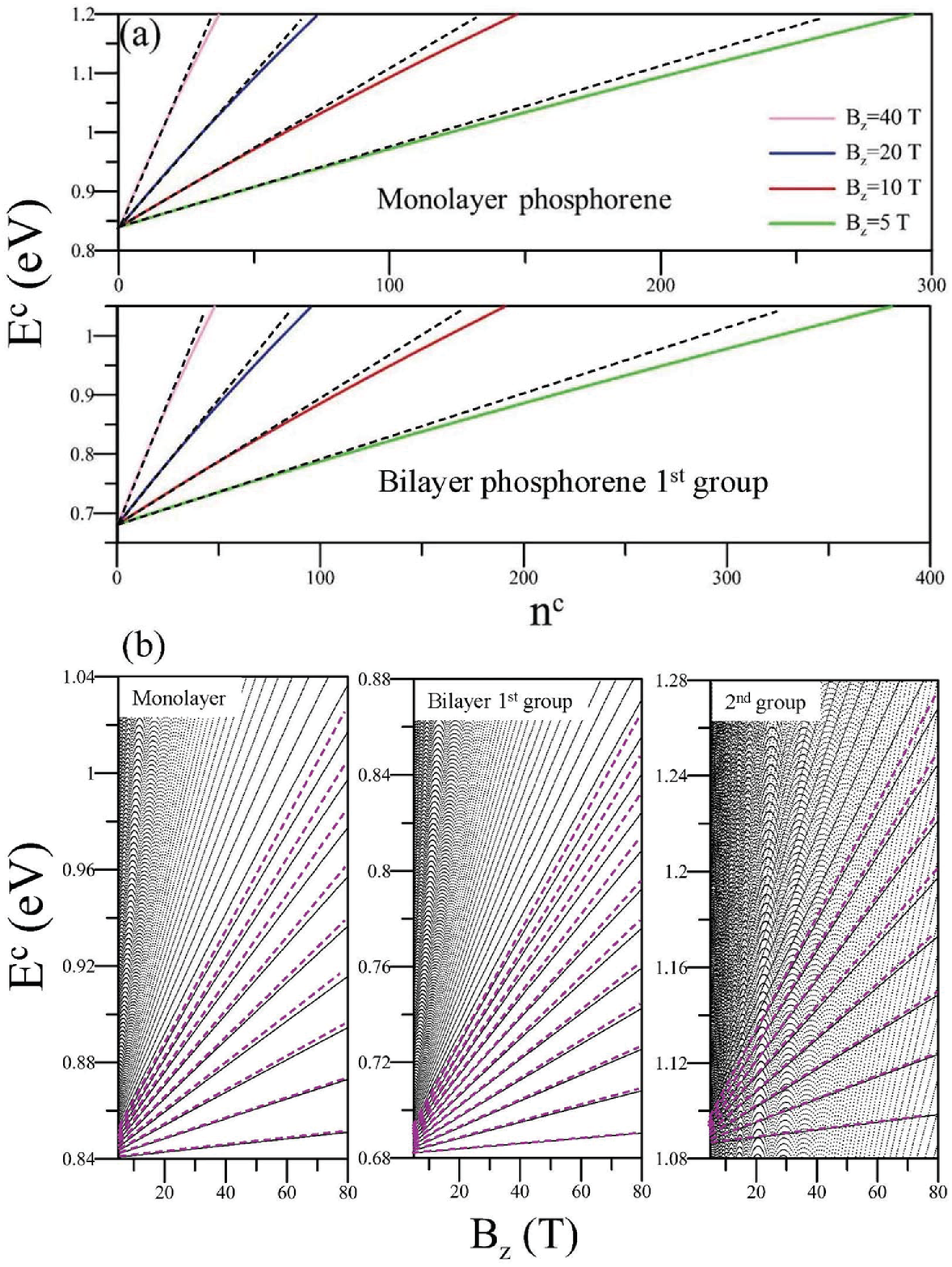}
\caption{The $n^v$-dependent LL energy spectra of monolayer and bilayer phosphorenes at various $B_z$’s, and (b) for the $B_z$-dependent ones.
}
\label{FIG:3}
\end{figure}

\begin{figure}
\centering
\includegraphics[width=1\textwidth]{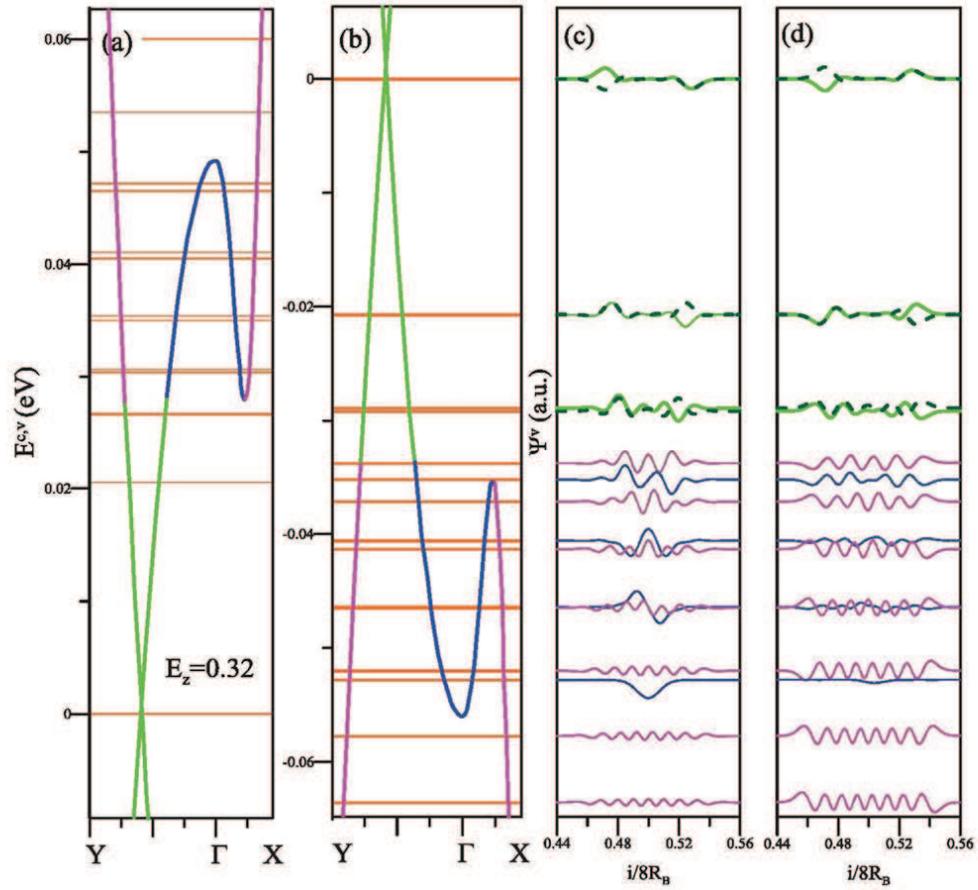}
\caption{At ${E_z=0.32}$, (a) the conduction and (b) valence LLs, and
the subenvelope functions of the latter on the (c) first and (d) second layers.}
\label{FIG:4}
\end{figure}

\begin{figure}
\centering
\includegraphics[width=1\textwidth]{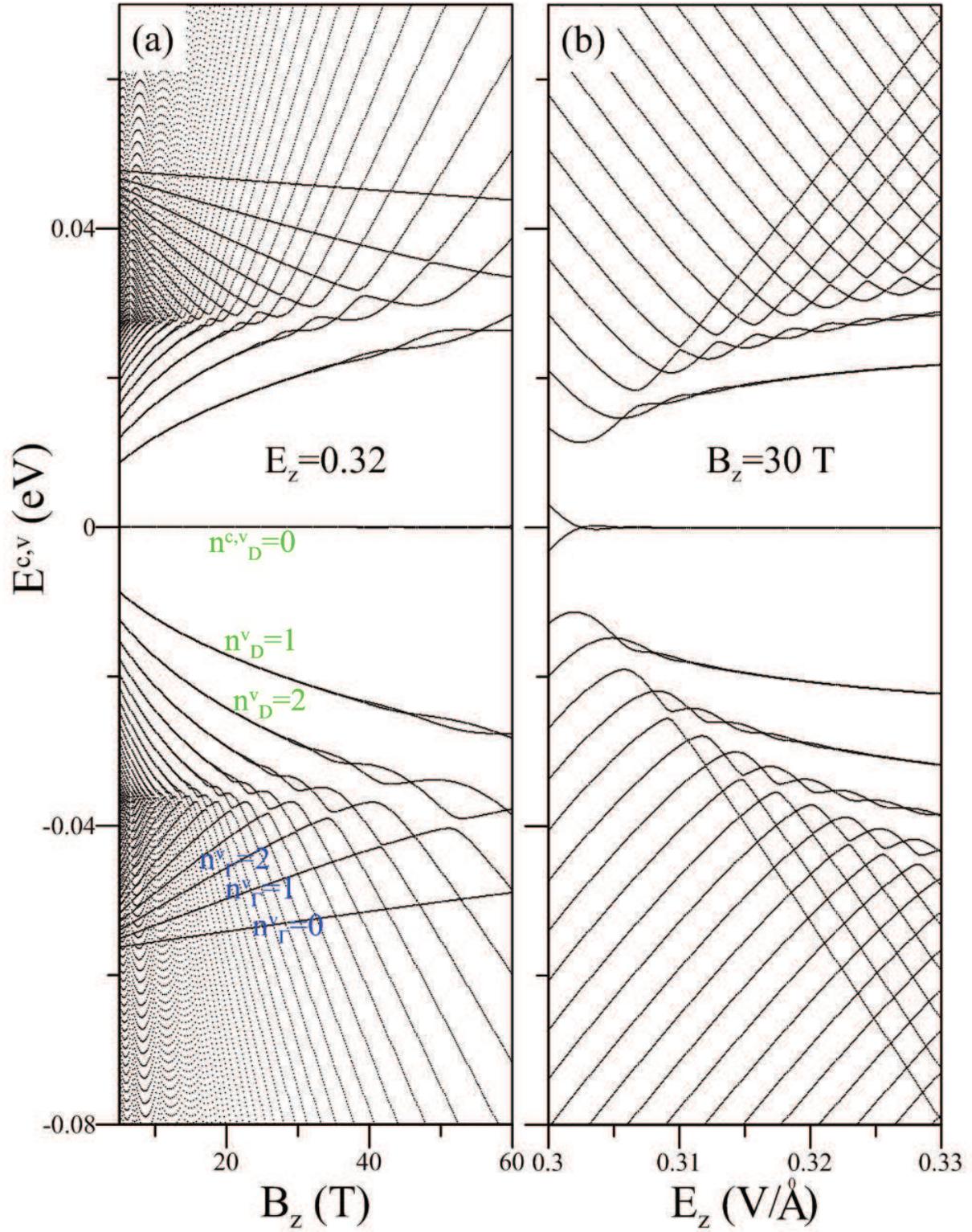}
\caption{(a) The $B_z$- and (b) $E_z$-dependent LL energy spectra.}
\label{FIG:5}
\end{figure}

\begin{figure}
\centering
\includegraphics[width=1\textwidth]{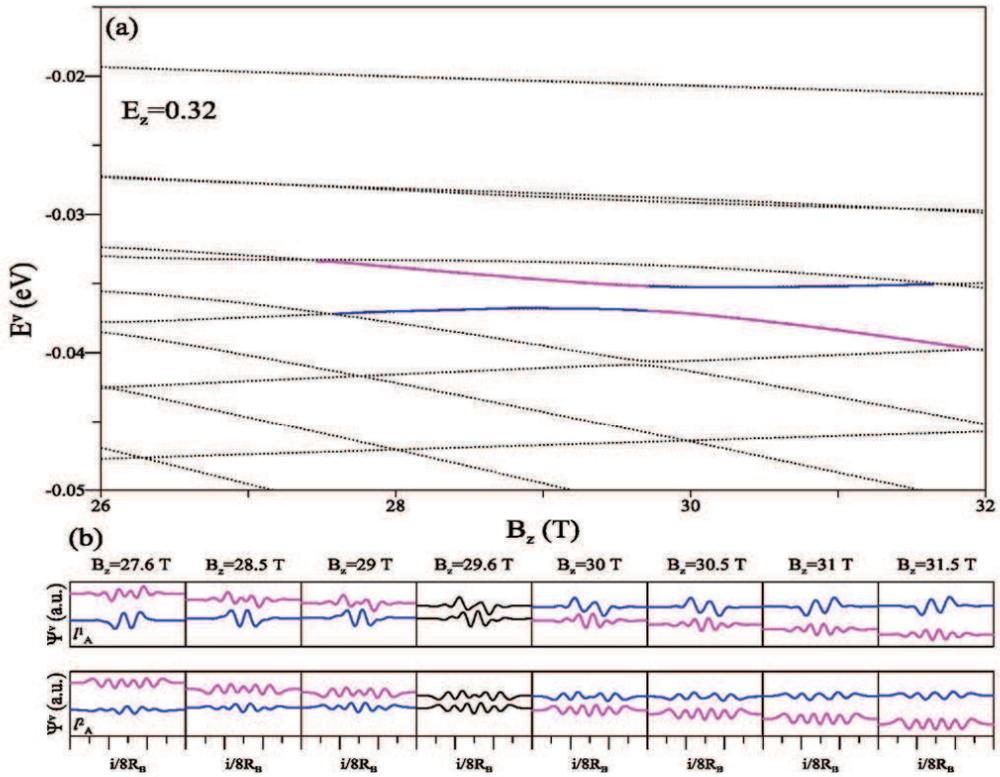}
\caption{(a) The LL anti-crossing energy spectrum (blue and pink curves), and (b) the dramatic transformation between oscillation modes on the first and second layers.}
\label{FIG:6}
\end{figure}

\begin{figure}
\centering
\includegraphics[width=1\textwidth]{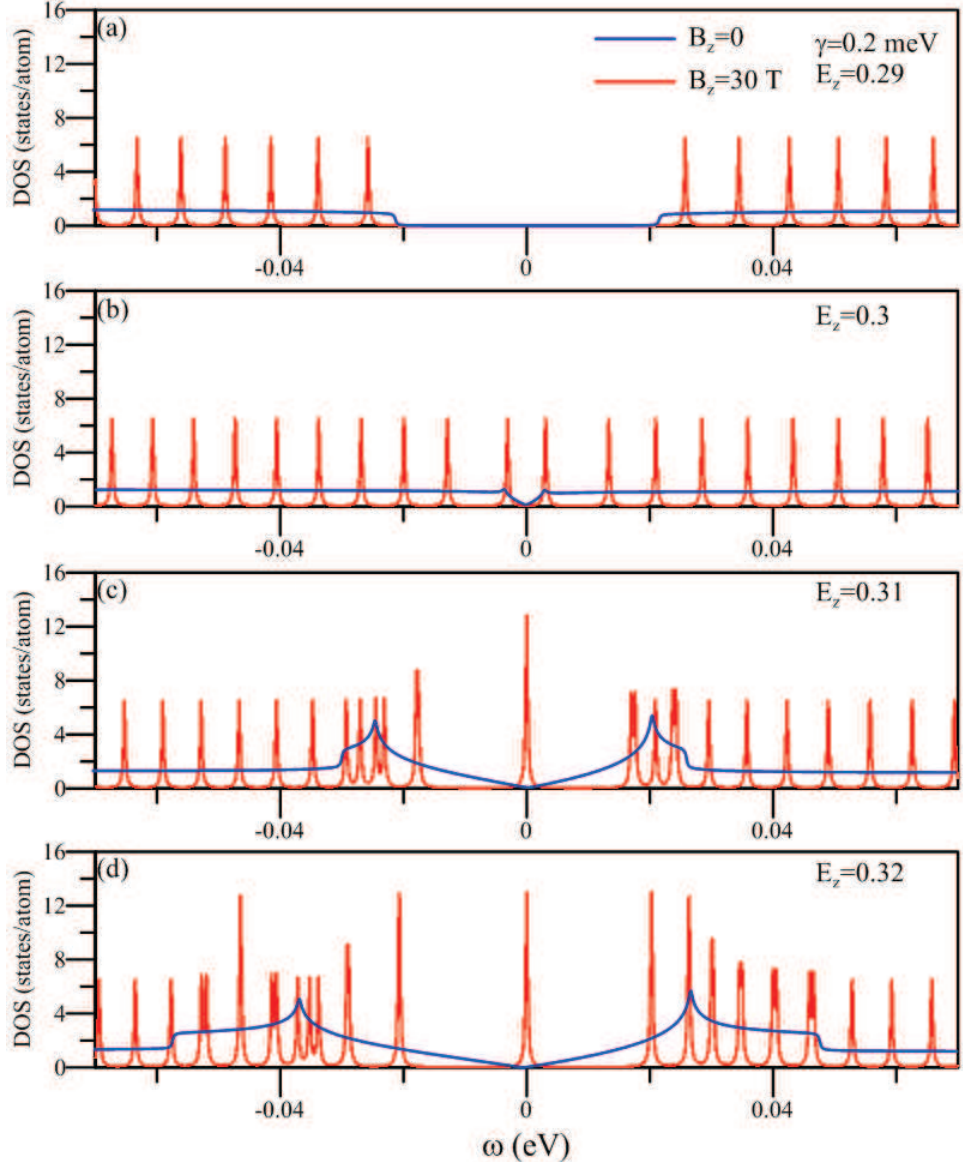}
\caption{DOSs of bilayer system at $B_z$=0 and 30 T (blue and red curves) (a)-(d) under various $E_z$’s. $\gamma$=0.2 meV is the broadening factor in calculating DOS.}
\label{FIG:7}
\end{figure}

\end{document}